\documentclass[amssymb,twocolumn,showpacs,amsmath,aps]{revtex4}
\usepackage{epsfig}

\begin{document}
 
\title{Influence of adatom interactions on second layer nucleation
       \vspace*{-2ex}}

\author{Stefan Heinrichs$^{1}$ and Philipp Maass$^{1,2}$}

\address{$^{1}$Fachbereich Physik, Universit\"at Konstanz, D-78457 Konstanz,
         Germany\\
         $^{2}$Fachbereich f\"ur Mathematik und Naturwissenschaften,
         Technische Universit\"at Ilmenau, 98684 Ilmenau, Germany}

\date{March 6, 2002}

\begin{abstract} 
  \vspace*{.4cm} We develop a theory for the inclusion of adatom
  interactions in second layer nucleation occurring in epitaxial
  growth. The interactions considered are due to ring barriers between
  pairs of adatoms and binding energies of unstable clusters.  The
  theory is based on a master equation, which describes the time
  development of microscopic states that are specified by cluster
  configurations on top of an island.  The transition rates are
  derived by scaling arguments and tested against kinetic Monte-Carlo
  simulations. As an application we reanalyze experiments to determine
  the step edge barrier for Ag/Pt(111).
\end{abstract}

\pacs{68.55.-a, 68.55.Ac, 81.15.-z, 81.15.Aa}
%68.55.-a   Thin film structure and morphology
%68.55.Ac   Nucleation and growth:  microscopic aspects
%81.15.Aa   Theory and models of film growth
%81.15.-z   Methods of deposition of films and coatings; 
%           film growth and epitaxy

\maketitle

Whether thin films in epitaxy become rough or grow smoothly layer by
layer depends on the onset of second layer nucleation on top of
islands in the first layer: If the rate $\Omega$ of this nucleation is
large, it is likely that mounds are formed before layer completion,
while small second layer nucleation rates favor layer by layer growth.
Detailed theories have been developed in the past
\cite{Tersoff/etal:1994,Rottler/Maass:1999,Heinrichs/etal:2000,Krug/etal:2000,Krug:2000}
to predict $\Omega$ in dependence on the island size $R$, the ratio
$\alpha_{\rm\scriptscriptstyle S}\equiv(\nu_{\rm\scriptscriptstyle S}/
\nu_{\rm t})\exp(-\Delta E_{\rm\scriptscriptstyle
  S}/k_{\rm\scriptscriptstyle B}T)$ of the hopping rates $\nu_{\rm
  t}\exp(-E_D/k_{\rm\scriptscriptstyle B}T)$ and
$\nu_{\rm\scriptscriptstyle S}\exp(-E_{\rm\scriptscriptstyle
  S}/k_{\rm\scriptscriptstyle B}T)$ on top of the island and over the
island edge, respectively, and the ratio $D/Fa^4$ of the adatom
diffusivity $D=(\nu_{\rm
  t}a^{2}/4)\exp(-E_{D}/k_{\rm\scriptscriptstyle B}T)$ and deposition
flux $F$; $a$ is the lattice constant. Studies so far focused on
adatoms that form stable nuclei once more than $i$ of them cluster
together, but otherwise do not interact.

In this Letter we show how adatom interactions strongly influence the
second layer nucleation. We consider two types of these interactions:
{\it (i)} Clusters of size smaller than $i\!+\!1$ are metastable,
i.e.\ the detachment of an adatom or the break up of clusters into two
pieces requires a dissociation energy $E_{\rm\scriptscriptstyle
  dis}\gtrsim k_BT$, and {\it (ii)} the approach of two atoms by
diffusion is hindered by additional barriers. As a generic model we
consider a barrier $E_{\rm\scriptscriptstyle ring}$ in form of a ring
with radius $\xi$ around each adatom.  The occurrence of such barriers
was explored by Fichthorn and Scheffler based on extensive density
functional calculations \cite{Fichthorn/Scheffler:2000}. Similar
interaction effects can be caused by Shockley surface states
\cite{Hyldgaard/Persson:2000}.

The aim of this Letter is to develop a theory based on rate equations
that allows us to take into account these interaction effects for
nucleation processes in confined geometry as encountered on top of a
growing island. As an application we quantify possible errors when
determining the step edge barrier $E_{\rm\scriptscriptstyle S}$
\cite{Ehrlich/Hudda:1966+Schwoebel:1969} in second layer nucleation
experiments under the assumption of non-interacting adatoms. For the
Ag/Pt(111) system we compare the results at different levels of
sophistication of the underlying theory for determining the step edge
barrier.

In order to take into account the adatom interactions, we refer to our
rate equation approach for second layer nucleation outlined in
\cite{Heinrichs/etal:2000}. In this approach one considers one compact
island with radius $R$ that evolves in time according to some growth
law $R=R(t)$. To clearly separate the different types of interaction
effects, we consider the presence of either metastable clusters or the
ring barrier.

We denote by $p_{n,\nu}(t)$ the probability to find the island in a
state where in total $n$ atoms are on top of the island in
configuration $\nu$ at time $t$. The configuration label $\nu$
specifies the way the $n$ atoms are decomposed into clusters and which
pairs of atoms have a distance smaller than $\xi$ and are thus
``weakly bound'' by their ring barriers.

%****************************************************************
\begin{figure}[t]
\begin{center}\epsfig{file=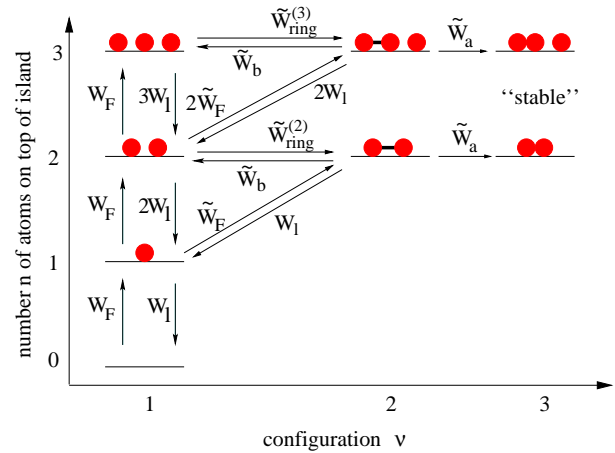,width=8cm}\end{center}
\vspace*{-0.2cm}
\caption{States and the corresponding transition rates
  involved in the second layer nucleation with ring barrier interaction and a
  critical nucleus of size $i=1$. The adatoms in configuration 2 are inside
  each others` ring barrier, which is marked by the connecting line. The two
  states with configuration label $\nu=3$ contain a stable dimer.}
\label{rate-illus-fig}
\end{figure}
%****************************************************************

In Fig. \ref{rate-illus-fig} the most important states are shown
together with the rates $W_{n,\nu\to n',\nu'}$ connecting them
\cite{states-comm}. The probabilities $p_{n,\nu}(t)$ obey the master
equation
\begin{equation}
\frac{dp_{n,\nu}}{dt}=\sum_{n',\nu'}
\bigl[W_{n',\nu'\to n,\nu}\,p_{n',\nu'}
-W_{n,\nu\to n',\nu'}\,p_{n,\nu}\bigr]\,,
\label{master-eq}
\end{equation}
where $W_{n,\nu\to n',\nu'}$ is the rate from state $(n,\nu)$ to
$(n',\nu')$ and can depend on time through the growth law $R=R(t)$
(see below). Equations (\ref{master-eq}) are numerically solved
subject to the initial condition $p_{n,\nu}\!=\!\delta_{n,0}$.

The rates $W_{n,\nu\to n',\nu'}$ are expressed in terms of elementary
rates $W_F$, $W_l$, $\tilde W_{\rm\scriptscriptstyle ring}^{(n)}$,
$\tilde W_F$, $\tilde W_b$ and $\tilde W_a$ (cf.
Fig.~\ref{rate-illus-fig}). The first two rates $W_F$ and $W_l$ refer
to processes not involving the ring barrier: $W_F=\pi FR^{2}$ is the
deposition rate of adatoms onto the island and
$W_l=(D/R^{2})[\kappa_{1} a/(\alpha_{\rm s} R) + \kappa_{2}]^{-1}$ is
the loss rate of adatoms that leave the island by surmounting the step
edge barrier \cite{Heinrichs/etal:2000}. The coefficients
$\kappa_{1},\kappa_{2}$ are of order one \cite{Heinrichs/etal:2000}.
In order to account for the reduced mobility of pairs of atoms bound
via the ring barrier compared to individual atoms, the loss rate for
such pairs is approximated to be equal to that of a single atom.

The latter four rates involve the ring barrier in form of the
associated Boltzmann factor $\alpha_{\rm\scriptscriptstyle
  ring}=\exp[-\Delta E_{\rm\scriptscriptstyle
  ring}/k_{\rm\scriptscriptstyle B}T]$, $\Delta
E_{\rm\scriptscriptstyle ring}=E_{\rm\scriptscriptstyle ring}-E_D$,
and the ring radius $\xi$.  $\tilde W_F=\pi F\xi^{2}$ is the
deposition (``flux'') rate into a ring. $\tilde W_a=\tilde \kappa_a
2D/(\pi\xi^2)$ is the attachment rate for two atoms in a circle which
are confined to a separation distance smaller than $\xi$. It results
from the time $\propto \xi^2/2D$ for the two atoms to encounter by
diffusion on top of an island with radius $\xi$. For the relevant case
$a<\xi\ll R$ we can further derive the rates for formation and breakup
of pairs weakly bound by their ring barriers. The breakup rate is
given by
\begin{equation}
\tilde W_b=\tilde
%\kappa_{b}(D\alpha_{{\rm\scriptscriptstyle ring}})/(a\xi) 
\kappa_{b}\frac{D\alpha_{{\rm\scriptscriptstyle ring}}}{a\xi}.
\label{Wb-eq}
\end{equation}
It results from the probability $\propto 2\pi \xi/\pi \xi^{2}$ for two
atoms to have distance $\xi$ times the rate $\propto
\alpha_{\rm\scriptscriptstyle ring}D/a^{2}$ to overcome the ring
barrier. The rate for formation of an adatom pair weakly bound by
their ring barriers is
\begin{equation}
\tilde W_{\rm\scriptscriptstyle ring}^{(n)}=
\frac{n(n\!-\!1)D}{R^2} \biggl(\tilde
  \kappa_{1}\frac{a}{\xi\alpha_{\rm\scriptscriptstyle ring}}+\tilde
  \kappa_{2}\biggr)^{-1},
\label{Wint-eq}
\end{equation}
if in total $n$ adatoms are on top of the island.  The associated time
$(\tilde W_{\rm\scriptscriptstyle ring}^{(2)})^{-1}$ for two atoms
results from two contributions: The time $\propto R^{2}/D$ for first
reaching the interaction distance $\xi\ll R$ and the time for
overcoming the barrier. The latter is given by $[(2\pi \xi a/\pi R^2)
\times (\alpha_{{\rm\scriptscriptstyle ring}}D/a^{2})]^{-1}$, where
the term $2\pi \xi a/\pi R^2$ is due to the probability for a pair to
have a distance in the interval $(\xi,\xi+a)$ and the term
$\alpha_{{\rm\scriptscriptstyle ring}}D/a^{2}$ is the rate to overcome
the ring barrier. The rate $\tilde W_{\rm\scriptscriptstyle
  ring}^{(2)}$ for a single pair has to be multiplied by the number
$n(n-1)/2$ of distinct pairs to obtain the total rate for $n$ adatoms.

%****************************************************************
\begin{figure}[t!]
\begin{center}\epsfig{file=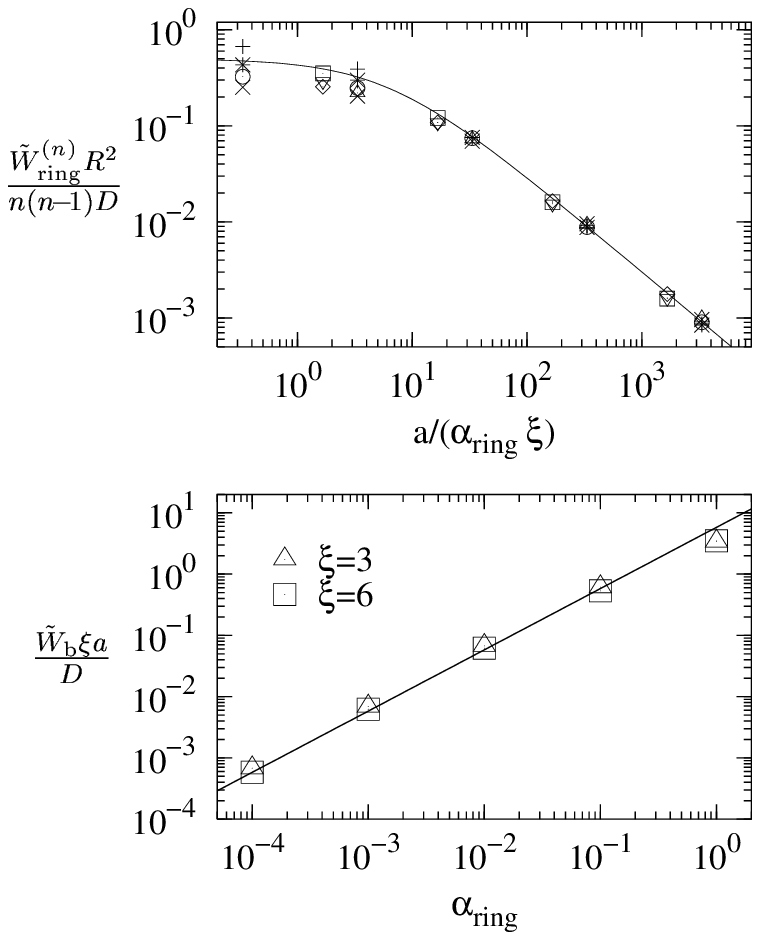,width=8cm}\end{center}
\vspace*{-0.2cm}
\caption{Scaling plots of the rates for formation and breakup of atom
  pairs bound by their ring barriers. The symbols in the upper part of
  the figure refer {\it (i)} to $n=2$, $\xi/a=3$ for $R/a=10$ ($+$),
  22 ($\ast$), 46 ($\circ$), 100 ($\times$), {\it (ii)} to $n=2$,
  $\xi/a=6$ for $R/a=46$ ($\scriptscriptstyle\square$), 100 ($\diamond$), and {\it (iii)}
  to $n=3$, $R/a=6$ for $\xi/a=3$ ($\scriptscriptstyle\triangle$) and 6
  ($\triangledown$). The solid line in the upper part is a fit to
  $f(x)=(\tilde\kappa_1 x+\tilde\kappa_2)^{-1}$ [cf.\ 
  eq.~(\ref{Wint-eq})] and the solid line in the lower part has slope
  one [cf.\ eq.~(\ref{Wb-eq})].}
\label{kmc-fig}
\end{figure}
%****************************************************************

The validity of all formulae for the elementary rates is tested
against kinetic Monte-Carlo (KMC) simulations performed on a hexagonal
lattice for the (111) surface. As an example we show in
Fig.~\ref{kmc-fig} scaling plots for the two rates $\tilde
W_{\rm\scriptscriptstyle ring}^{(n)}$ and $\tilde W_b$ demonstrating
the good agreement of the KMC data with the predictions of
eqs.~(\ref{Wb-eq},\ref{Wint-eq}). All constants $\tilde\kappa$ in
Eqs.~(\ref{Wb-eq},\ref{Wint-eq}) are of the order of one
\cite{kappa-comm}. The approach outlined above provides a general
framework to treat the problem of second layer nucleation and may be
applied and extended to a variety of situations. It furthermore allows
one to gain detailed insight into the microscopic pathways followed
during the nucleation process.

To apply our theory we consider the determination of step edge barriers in
second layer nucleation experiments. In these experiments one measures the
probability $f[R(t)]$ that a stable cluster has nucleated on top of the island
until time $t$ (i.e.\ the fraction of ``covered islands''). In systems with a
substrate mediated ring barrier, second layer nucleation will be aggravated
and nucleation sets in later. A repulsive ring barrier therefore has an effect
similar to a reduced step edge barrier, and thus yields an apparent
measurement value $\Delta E_{\rm S}^{(0)}$ smaller than the ``true'' $\Delta
E_{\rm S}$. On the other hand, metastable dimers facilitate the formation of a
stable nucleus, leading to values $\Delta E_{\rm S}^{(0)}$ larger than $\Delta
E_{\rm S}$.

\begin{figure}[t]
\epsfig{file=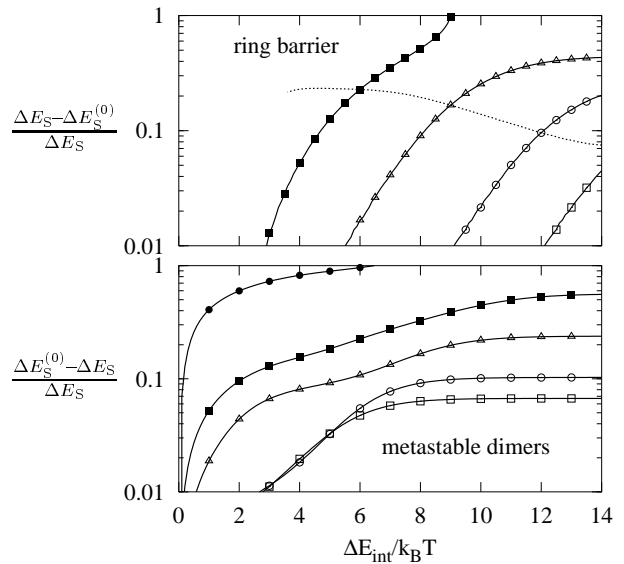,width=8cm}
\vspace*{0.2cm}
\caption{Relative error of the apparent additional step edge barrier $\Delta
  E_{\scriptscriptstyle\rm S}^{(0)}$ when neglecting interactions ($\Delta
  E_{\rm int}=0$) with respect to the ``true'' barrier $\Delta
  E_{\scriptscriptstyle\rm S}$ calculated for $\Delta E_{\rm int}>0$; upper
  part: $\Delta E_{\rm int}=\Delta E_{{\rm\scriptscriptstyle ring}}$, lower
  part: $\Delta E_{\rm int}=\Delta E_{{\rm\scriptscriptstyle dis}}$. The
  symbols refer to different values $\Delta E_{\rm\scriptscriptstyle
  S}/k_{\rm\scriptscriptstyle B}T=3$
  ($\bullet$), 6 ($\scriptscriptstyle\blacksquare$), 9
  ($\scriptscriptstyle\triangle$), 12 ($\circ$), and 15
  ($\scriptscriptstyle\square$).  The dotted line in the upper figure is drawn
  for $\Delta E_{\scriptscriptstyle\rm S}=\Delta E_{{\rm\scriptscriptstyle
      ring}}$.  All plots are calculated with $D/Fa^{4}=10^{9}$ and the
  generic growth law $R(t)\propto\sqrt{t}$.}
\label{interaction-fig}
\end{figure}

To estimate the significance of the interactions, we first present results for
the relative error $[\Delta E_{\rm S}-\Delta E_{\rm S}^{(0)}]/\Delta E_{\rm
  S}$ as a function of the interaction parameters. By solving
Eq.~(\ref{master-eq}) we obtain the curves $f[R(t);\Delta
E_{\scriptscriptstyle\rm S}]$ that refer to processes including interactions.
On the other hand, we can solve the rate equations neglecting the interaction
($\Delta E_{{\rm\scriptscriptstyle ring}}=0$), thus obtaining
$f_{0}[R(t);\Delta E_{\scriptscriptstyle\rm S}^{(0)}]$. By fitting the curves
$f_{0}$ to the ``true '' curves $f$ we obtain the
apparent $\Delta E_{\scriptscriptstyle\rm S}^{(0)}$. For the ratio of attempt
frequencies $\nu_{\scriptscriptstyle\rm S}/\nu_{\rm t}$ for
adatom hopping over the step edge and on a terrace, we use the generic value
of one (for a general discussion on attempt frequencies, see
\cite{Ovesson/etal:2001}).  Using a generic growth law $R(t)\propto \sqrt{t}$
(with the prefactor $(D/Fa^{4})^{i/2(i\!+\!2)} \sqrt{Fa^{2}}$ from standard
nucleation theory \cite{Venables/etal:1984}) we show in the upper half of
Fig.~\ref{interaction-fig} results for the relative error as a function of
$\Delta E_{\rm\scriptscriptstyle ring}$. In the lower half of the figure we
include the results of an analogous analysis for $i=2$ with metastable dimers
(for a setup of the corresponding rate equations, see
\cite{Heinrichs/etal:2000}). Because the dependence on $D/Fa^{4}$ is only
weak, this parameter is not varied in the plots. We see that for small $\Delta
E_{\scriptscriptstyle\rm S}$, even weak interactions $\Delta E_{\rm int}$ lead
to a large relative error.  Closer inspection shows, that for fixed ratio
$\Delta E_{\rm int}/\Delta E_{\scriptscriptstyle\rm S}$, the relative error
decreases with increasing interaction energy (cf.\ the dashed line in
Fig.~\ref{interaction-fig}). In the case of metastable dimers, the error
reaches a plateau when the dissociation energy becomes so large that a dimer
is stable on the time scale of the formation of a stable trimer. This
signifies the transition to the $i\!=\!1$ case.

%****************************************************************
\noindent
%****************************************************************

For Ag/Pt(111) detailed second layer nucleation measurements of the type
discussed above were performed by Bromann {\it et al.}
\cite{Bromann/etal:1995}. The system Ag/Pt(111) is particularly suited as a
reference, since it has the advantage that many of the relevant parameters
were determined both by experiment
\cite{Brune/etal:1995,Bromann/etal:1995,Bott/etal:1996,Brune/etal:1999} and by
first-principle calculations
\cite{Feibelman:1994,Ratsch/etal:1997,Ratsch/Scheffler:1998,Fichthorn/Scheffler:2000}.
Experimentally, the diffusion barrier $E_D$ for Ag adatom diffusion on
strained Ag islands grown on Pt(111) is $E_D\cong 60$~meV
\cite{Brune/etal:1995}.  It has been shown by density functional calculations
\cite{Fichthorn/Scheffler:2000} that silver adatoms diffusing on top of an
already existing silver island on the platinum surface exhibit a strong
barrier $\Delta E_{\rm ring}=E_{\rm ring}-E_D\cong50$~meV at a distance
$\xi/a\cong 2.1$.

\begin{figure}[t!]
\begin{center}\epsfig{file=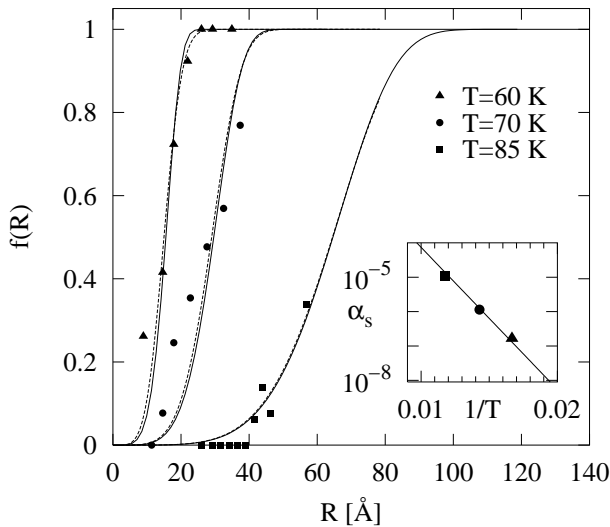,width=8cm}\end{center}
\vspace*{-0.2cm}
\caption{Fraction of covered Ag islands of size $R$ 
  as measured by Bromann {\it et al}.\ \cite{Bromann/etal:1995}. The solid
  lines refer to fits with the scaling approach when neglecting
  interactions and the dashed lines refer to fits with the rate equation theory
  when including the ring barriers.  The inset shows the Arrhenius fit for
  values $\alpha_{\rm s}$ obtained by the rate equation theory for
  $\nu_{\rm\scriptscriptstyle S}=\nu_{\rm t}=10^{12}$~Hz.}
\label{bromann-fig}
\end{figure}

The original analysis by Bromann {\it et al}.\ was based on a mean
field-type theory for second layer nucleation (``TDT approach'', see
\cite{Tersoff/etal:1994}). However, it has been shown recently that
even in the absence of interaction effects it is necessary to
re-analyze the data with an extended theory
\cite{Rottler/Maass:1999,Heinrichs/etal:2000} (see also
\cite{Krug/etal:2000}). In this theory fluctuation-dominated regimes
occur for critical nuclei $i\le2$ \cite{Heinrichs/etal:2000}, and
it yields, for the experimental conditions and a prefactor $\nu_{\rm t}=10^{9}$~Hz
used in \cite{Bromann/etal:1995}, the second-layer nucleation rate
$\Omega(R)=\pi^{2} \kappa_{1} F R^5 /(\Gamma \alpha_{\rm s})$ where
$\Gamma=D/Fa^4$. By contrast to the
generic growth law $R(t)\propto \sqrt{t}$ used for
Fig.~\ref{interaction-fig}, in the analysis of Bromann {\it et al.}
one has to deal with an exponential growth law $R(t)$ (for details see
\cite{Bromann/etal:1995}). Metastable states are not considered here, because $i=1$ in the temperature range
of the experiment.  In Fig.~\ref{bromann-fig} we have fitted
the measured fraction of covered islands $f(R)$ yielding $\Delta
E_{\rm\scriptscriptstyle S}^{(0)}=65$~meV, which is about twice the value
obtained previously based on the TDT approach (see Table~I). For the
much less reliable prefactor ratio \cite{Roos/Tringides:2000} we find
$\nu_{\rm\scriptscriptstyle S}/\nu_{\rm t}=41$. The rate equation theory without interactions yields essentially the
same value $\Delta E_{\rm\scriptscriptstyle S}^{(0)}=68$~meV thereby
supporting the scaling approach
\cite{Rottler/Maass:1999,Heinrichs/etal:2000}.

%****************************************************************
%****************************************************************

The next step in theoretical sophistication is the inclusion of the ring
barrier. When using the value $\Delta E_{\rm\scriptscriptstyle ring}=50$~meV
\cite{Fichthorn/Scheffler:2000}, we find a significantly decreased value of
$\Delta E_{\rm\scriptscriptstyle S}=52$~meV. This decrease seems to be in
contradiction to the reasoning above that $\Delta E_{\rm\scriptscriptstyle
  S}^{(0)}<\Delta E_{\rm\scriptscriptstyle S}$. However, the values were
obtained by fitting both the prefactor $\nu_{\rm\scriptscriptstyle
  S}/\nu_{\rm t}$ and $\Delta E_{\rm\scriptscriptstyle S}$
to the data and cannot be directly compared because the prefactors turn out to
be very different (see table~I). When fitting under the constraint
$\nu_{\rm\scriptscriptstyle S}/\nu_{\rm t}=1$
\cite{Krug/etal:2000}, the value for $\Delta E_{\rm\scriptscriptstyle S}$
increases from $42$~meV to $48$~meV in accordance with the general argument.
Interestingly, the behavior of the standard error (see table~I) indicates that
the system is better described when including the interactions. However, this
interpretation should not be driven too far, since the standard error is only
an indication of the quality of the fit, but should not be interpreted as the
measurement error.

%****************************************************************
\begin{table}[h]
\begin{tabular}{crc|cc|c}
\begin{minipage}[c]{.7cm} $\nu_{\rm t}$ [Hz]
\end{minipage}
& \begin{minipage}[c]{1.2cm} $\Delta E_{\scriptscriptstyle\rm ring}$ [meV]
\end{minipage}
& Theory &
\begin{minipage}[c]{1.2cm}
$\Delta
E_{\scriptscriptstyle\rm S}$ [meV]
\end{minipage}
&$\nu_{\rm\scriptscriptstyle S}/\nu_{\rm t}$&
\begin{minipage}[c]{2cm} $\Delta E_{\scriptscriptstyle\rm S}$ [meV]
  $(\nu_{\rm\scriptscriptstyle S}/\nu_{\rm t}\!=\!1)$
\end{minipage}
\\[.2cm]
\hline
$10^9$ & 0 \hspace*{.5cm}& TDT & $30\pm5$ & 1& $30$\\
$10^9$ & 0 \hspace*{.5cm}& Scaling & $65\pm5$& 41& $43\pm2.2$\\
$10^9$ & 0 \hspace*{.5cm}& Rate Eq. & $68\pm1$ & 83 & $42\pm2.6$ \\
$10^9$&50  \hspace*{.5cm}& Rate Eq. & $52\pm1$ &2 & $48\pm0.4$ \\
$10^{12}$&50 \hspace*{.5cm}& Rate Eq. & $74\pm2$ &0.3 & $82\pm0.8$ \\
\end{tabular}
\label{comp-table}
\caption{Comparison of results for the additional step edge barrier for
  Ag/1ML Ag/Pt(111) obtained with theories of different levels 
of sophistication (see text). The standard errors from the fitting
procedure are shown for the values of $\Delta
E_{\scriptscriptstyle\rm S}$. }
\end{table}
%****************************************************************

It was recently argued that the often reported small prefactors for
hopping on weakly corrugated surfaces are due to neglecting the
interaction effects in the analysis of first layer nucleation
experiments \cite{Ovesson/etal:2001}. We therefore also analyze the
data using the theoretical prediction $\nu_{\rm t}=10^{12}$~Hz
\cite{Ratsch/Scheffler:1998} in the rate equation approach. The three
$\alpha_{\rm\scriptscriptstyle S}$ values obtained for each
temperature shown in Fig.~\ref{bromann-fig} lie very closely on a line
in an Arrhenius plot (see the inset in Fig.~\ref{bromann-fig}).
Both when using $\nu_{\scriptscriptstyle\rm S}/\nu_{\rm t}$ as
fitting paramter and when setting 
$\nu_{\scriptscriptstyle\rm S}/\nu_{\rm t}=1$, we find comparatively
large values $\Delta E_{\rm\scriptscriptstyle S}=74$~meV and
$\Delta E_{\rm\scriptscriptstyle S}=82$~meV, respectively.

In summary we have developed a rate equation theory that allows us to
quantify the influence of certain interaction effects on the
nucleation on top of islands in epitaxial growth. The formalism is
rather general and may be extended to other types of adatom
interactions and other kinds of geometrical constraints. We applied
the theory to second layer nucleation experiments designed to extract
the additional step edge barrier. It was shown that the value of
$\Delta E_{\rm\scriptscriptstyle S}$ depends very sensitively on
whether the interactions are carefully accounted for in the analysis.

We are indebted to W.~Dieterich and F.~Scheffler for very helpful
discussions and thank the Deutsche Forschungsgemeinschaft for
financial support through the SFB 513 and the Heisenberg program
(P.M., Ma 1636/2-1).


\begin{references}
\bibitem{states-comm} We have checked that inclusion of further states in the
  analysis does not change our results by more than a few percent except for
  very large ratios $\Delta E_{\rm\scriptscriptstyle
    S}/k_{\rm\scriptscriptstyle B}T$ and $D/Fa^4\lesssim10^{5}$.
\bibitem{kappa-comm} The results for breakup and formation of ring
  barriers from the
  KMC simulations were renormalized so that in the limit of weak
  barriers the nucleation rate without ring barriers is approximately
  reproduced. The values used are: $\tilde\kappa_1=.34$,
  $\tilde\kappa_2=1$, $\tilde\kappa_{\rm b}=0.5$ and
  $\tilde\kappa_{\rm a}=1$.
\end{references}
\end{document}